\documentstyle[seceq,epsf,preprint]{ptptex}

\newcommand{\bra}[1]{\langle {#1} |}     
\newcommand{\ket}[1]{| {#1} \rangle}     
\newcommand{\bbra}[1]{\langle\!\langle {#1} |}     
\newcommand{\kket}[1]{| {#1} \rangle\!\rangle}     
\newcommand{\rket}[1]{| {#1} )}     
\newcommand{\wtilde}[1]{\widetilde{#1}} 
\newcommand{\ol}[1]{\overline{#1}} 

\title{
The Lipkin Model in the $su(M+1)$-Algebra for\\
Many-Fermion System and its Counterpart in the\\
Schwinger Boson Representation}

\author{
Constan\c{c}a {\sc Provid\^encia},$^{1}$ 
Jo\~ao da {\sc Provid\^encia},$^{1}$ Yasuhiko {\sc Tsue}$^{2}$ 
and Masatoshi {\sc Yamamura}$^{3}$
}

\inst{
$^{1}$Departamento de Fisica, Universidade de Coimbra, P-3000 Coimbra, 
Portugal\\
$^{2}$Physics Division, Faculty of Science, Kochi University, Kochi 780-8520, 
Japan\\
$^{3}$Faculty of Engineering, Kansai University, Suita 564-8680, Japan
}

\recdate{
\today
}

\abst{
Following the Schwinger boson representation for the $su(M+1)$- and the 
$su(N,1)$-algebra presented by two of the present authors (J. da P. and M. Y.) 
and Kuriyama, a possible counterpart of the Lipkin model in 
the $su(M+1)$-algebra formulated in the fermion space is presented. 
The free vacuum, which plays a fundamental role in the conventional treatment 
of the Lipkin model, is generalized in a quite natural way, and further, 
the excited state generating operators such as the particle-hole pairs 
are also given in a natural scheme. 
As concrete examples, the cases of the $su(2)$-, $su(3)$- and the 
$su(4)$-algebra are discussed. 
Especially, the case of the $su(4)$-algebra is investigated in detail in 
relation to the nucleon pairing correlations and the high temperature 
superconductivity. 
}

\begin{document}

\maketitle

\section{Introduction}

The Lipkin model,\cite{1} which was proposed by Lipkin, Meshkov and Glick 
in 1965, has played a crucial role in microscopic studies of collective 
motions observed in nuclei. This is one kind of shell models: 
Under a certain interaction, many fermions move in two single-particle 
levels with the same degeneracy. Also, this is one kind of the 
$su(2)$-algebraic models: 
The Hamiltonian is expressed in terms of the $su(2)$-generators. 
However, implicitly the degrees of freedom, which do not connect 
directly with the algebra, are contained. 
This model has contributed to the schematic understanding of collective motion 
induced by certain type of the particle-hole pair excitations in 
the closed shell nuclei, i.e., fermion number of the system under 
investigation is equal to the degeneracy of the single-particle level. 
Then, as for the uncorrelated ground state called the free vacuum, we 
can adopt a state, in which the lower single-particle level is occupied 
fully by fermions. 
This model contains only one excited state generating operator 
from the free vacuum, if we restrict ourselves to the framework of 
the $su(2)$-algebra. 
With the use of this operator, we can describe the particle-hole pair 
excitation. 
Therefore, with the help of this model, only one kind of collective 
motion can be treated, and naturally, it is impossible to investigate 
the coupling to the other degrees of freedom in the framework of the 
$su(2)$-algebra. 
A possible idea to answer to the above mentioned problem may be to 
generalize the case of two single-particle levels to the case of three 
with the same degeneracy, i.e., the $su(3)$-algebraic model.\cite{2} 
Depending on the fermion number, this model has two closed-shell states, i.e., 
two free vacuums. 
One is the state, in which the lowest single-particle level is fully occupied 
by fermions, and the other is the state, in which the lower two 
single-particle levels are fully occupied. 
The particle-hole pair Tamm-Dankoff approximation suggests us 
that in this model there exist two excited state generating operators for 
each vacuum. 
Therefore, we can investigate the coupling between two kinds of the 
excitations. 
If, in the framework of these two generating operators, collective and 
non-collective degrees of freedom are defined, we can describe the coupling 
of the collective motion to the non-collective one. 
This is one of the fundamental problem in nuclear theory. 
It is the reason why the generalization of the $su(2)$- to the $su(3)$-model 
is necessary.

Under the above-mentioned situation, the generalization to the 
$su(4)$-algebraic model is meaningful or not ? 
It may be not necessary, if we adhere to the same viewpoint as that in the 
$su(3)$-model, i.e., the free vacuum is given only in the form that 
the lowest level is fully occupied and the others are vacant. 
However, if we are free from the above adherence, the reply is that 
we are compelled to generalize. 
This task is performed by introducing four single-particle levels 
with the same degeneracy. 
Depending on the fermion number, this model contains three 
closed-shell states, i.e., three free vacuum: 
(i) The lowest single-particle level is fully occupied, (ii) the 
lower three single-particle levels are fully occupied and 
(iii) the lower two levels are fully occupied. 
The particle-hole pair Tamm-Dankoff approximation suggests us that, 
in the cases (i) and (ii), there exist three excited state generating 
operators, but in the case (iii), there exist four. 
The above means that depending on the difference of the free vacuums, 
the number of the excited state generating operators is different from 
each other. 
Therefore, we should clarify the meaning of this difference. 
Further, we know two $su(4)$-algebraic models in many-fermion systems. 
One is related to the isospin vector and scalar pairing correlations.\cite{3} 
This is a form extended from the $so(5)$-model for the isospin vector 
pairing correlation.\cite{4} 
The other is related to the high-temperature superconductivity,\cite{5} 
which was also extended from the $so(5)$-model.\cite{6} 
For investigating these $su(4)$-models, the use of the free vacuum, 
in which the lowest level is only occupied, seems to be helpless. 
In the above-mentioned reasons, it may be interesting to formulate the 
Lipkin model in the $su(M+1)$-algebra. 
Of course, this model consists of $(M+1)$ single-particle levels 
with the same degeneracy, in which each level is specified by 
$i=0,1,2,\cdots , M$. 
As a vacuum, the single-particle levels $i=0,1,2,\cdots ,l$ are occupied 
by fermions. 
Then, our problem is reduced to determine the number of the excited 
state generating operators, and further, the concrete forms of these 
operators.

Usually, the $su(2)$-, the $su(3)$-algebraic models and their generalization 
have been discussed in relation to the excitations from the free vacuum, 
in which the lower single-particle levels are fully occupied. 
However, the $su(2)$- and the $su(3)$-models have also contributed to the 
schematic understanding of finite temperature effects in many-fermion 
systems, for example, we can find the case of the $su(2)$-model in 
Ref. \citen{7} and the case of the $su(3)$-model in Ref. \citen{8}. 
Therefore, it may be a meaningful problem for the study of finite 
temperature effects to generalize the $su(2)$- and the $su(3)$-algebraic 
models. 
However, in this case, the use of the free vacuum is helpless even the 
closed-shell systems. 
For this purpose, the structure of the free vacuum must be extended. 
This task is also inevitable for the generalization. 
We will call the state extended from the free vacuum as the intrinsic state. 
In \S 3, the reason will be mentioned. 
We will name the generalization as the Lipkin model in the 
$su(M+1)$-algebra.

In addition to the above, it is well known that the boson expansion theories 
were proposed by Belyaev and Zelevinsky in 1962\cite{9} and in 
slightly different viewpoint by Marumori, Yamamura (one of the present 
authors) and Tokunaga in 1964.\cite{10} 
Afterward, the boson realization of the Lie algebra has played a central role 
in the schematic studies of nuclear many-body problem, and in particular, 
in the field of microscopic studies of collective dynamics. 
Concerning these studies, Ref. \citen{11} is instructive. 
In response to this circumstance, two of the present authors (J. da P. and 
M. Y.) and Kuriyama proposed a possible boson realization of the 
$su(M+1)$-algebra and its associated $su(N,1)$-algebra.\cite{12} 
Judging from the form analogous to the boson realization of the 
$su(2)$-algebra presented by Schwinger,\cite{13} 
we can call it the Schwinger boson representation. 
Further, this representation was applied to the Lipkin model in the 
$su(M+1)$-algebra and various interesting aspects of this model were 
investigated.\cite{14} 
However, the subjects, which were mentioned in this section, are not able to 
be solved transparently. 
The reason will be given in \S 5.

A main aim of this paper is to formulate a possible counterpart of the 
Lipkin model in the $su(M+1)$-algebra in the framework of the Schwinger 
boson representation developed in Ref. \citen{12}. 
As was already mentioned, even if we start in the free vacuum as a kind of 
the intrinsic states, this model suggests various interesting features 
of the dynamics of many-fermion systems. 
However, there exist the cases in which the free vacuum is helpless. 
For such cases, we must construct the intrinsic state suitable for 
each problem. 
But, for this construction, the degrees of freedom, which do not connect 
directly with the $su(M+1)$-algebra, appear explicitly. 
Of course, this discussion presupposes that the Hamiltonian and 
other quantities are expressed in terms of the generators of the algebra. 
From this reason, it may be a quite tedious task to construct the 
intrinsic state explicitly in the original fermion space. 
Also, the construction of the excited state generating operators 
contains also the same trouble as that in the above. 
In contrast to the above-mentioned situation, in the Schwinger boson 
representation, the system under investigation is expressed in terms of 
the minimum number of the degrees of freedom which describes the algebra. 
In this sense, the construction of the intrinsic state in the boson space 
may be much easier than that in the original space. 
This is the reason why we intend to develop a possible counterpart of the 
Lipkin model in the $su(M+1)$-algebra in the framework of the boson space. 
Of course, not only for the investigation of collective motions but also 
for the study of finite temperature effects, the results, which can be 
derived in the boson space, should be expected to be equivalent to those, 
which may be derived in the fermion space.

After recapitulating the Schwinger boson representation for the 
$su(M+1)$-algebra and its associated $su(N,1)$-algebra in \S 2, 
the Lipkin model in the $su(M+1)$-algebra is summarized in \S 3. 
In \S 4, the Holstein-Primakoff boson representation for the 
$su(M+1)$-algebra is presented together with the limitation of the 
applicability. 
Section 5 is devoted to constructing the intrinsic states in the Schwinger 
boson representation. 
In \S 6, various properties of the intrinsic states are presented. 
Finally, in \S 7, as the concrete example, the cases of the $su(2)$- and 
the $su(3)$-algebra are discussed, and further, the $su(4)$-algebra is 
investigated in relation to the nucleon pairing correlations 
and the high-temperature superconductivity. 
In Appendix, applicability of the intrinsic state in the boson space to 
the finite temperature effects is presented.

\section{The $su(M+1)$-algebra and its associated $su(N,1)$-algebra and 
their Schwinger boson representation}

The generators of the $su(M+1)$-algebra $({\hat S}^i, {\hat S}_i, 
{\hat S}_i^j\ ;\ i,j=1,2,\cdots ,M)$ obey the following relations: 
\begin{subequations}\label{2-1}
\begin{eqnarray}
& &{\hat S}_i^*={\hat S}^i \ , \qquad {\hat S}_j^{i*}={\hat S}_i^j \ , 
\label{2-1a}\\
& &[\ {\hat S}^i \ , \ {\hat S}^j\ ]=0\ , \qquad 
[\ {\hat S}^i \ , \ {\hat S}_j\ ]={\hat S}_i^j \ , \nonumber\\
& &[\ {\hat S}_i^j \ , \ {\hat S}^k \ ]=\delta_{jk}{\hat S}^i 
+ \delta_{ij}{\hat S}^k \ , \nonumber\\
& &[\ {\hat S}_i^j \ , \ {\hat S}_l^k \ ]=\delta_{jl}{\hat S}_i^k 
- \delta_{ik}{\hat S}_l^j \ . 
\label{2-1b}
\end{eqnarray}
The Casimir operator ${\hat \Gamma}_{su(M+1)}$ is expressed in the form 
\begin{eqnarray}\label{2-1c}
{\hat \Gamma}_{su(M+1)}&=&2\left(\sum_{i=1}^M {\hat S}^i {\hat S}_i 
+\sum_{j>i}{\hat S}_j^i {\hat S}_i^j \right) \nonumber\\
& &+\sum_{i=1}^M ({\hat S}_i^i)^2-(M+1)^{-1}
\left(\sum_{i=1}^M{\hat S}_i^i\right)^2+\sum_{i=1}^M(M-2i){\hat S}_i^i \ . 
\end{eqnarray}
\end{subequations}
The $su(N,1)$-algebra is composed of the operators 
$({\hat T}^p, {\hat T}_p, {\hat T}_q^p\ ;\ p,q=1,2,\cdots ,N)$ 
satisfying the following relations: 
\begin{subequations}\label{2-2}
\begin{eqnarray}
& &{\hat T}_p^*={\hat T}^p \ , \qquad {\hat T}_p^{q*}={\hat T}_q^p \ , 
\label{2-2a}\\
& &[\ {\hat T}^p \ , \ {\hat T}^q\ ]=0\ , \qquad 
[\ {\hat T}^p \ , \ {\hat T}_q\ ]=-{\hat T}_q^p \ , \nonumber\\
& &[\ {\hat T}_q^p \ , \ {\hat T}^r \ ]=\delta_{qr}{\hat T}^p 
+ \delta_{pq}{\hat T}^r \ , \nonumber\\
& &[\ {\hat T}_q^p \ , \ {\hat T}_r^s \ ]=\delta_{qs}{\hat T}_r^p 
- \delta_{pr}{\hat T}_q^s \ . 
\label{2-2b}
\end{eqnarray}
The Casimir operator ${\hat \Gamma}_{su(N,1)}$ is given in the form 
\begin{eqnarray}\label{2-2c}
{\hat \Gamma}_{su(N,1)}&=&-2\left(\sum_{p=1}^N {\hat T}^p {\hat T}_p 
-\sum_{q>p}{\hat T}_p^q {\hat T}_q^p \right) \nonumber\\
& &+\sum_{p=1}^N ({\hat T}_p^p)^2-(N+1)^{-1}
\left(\sum_{p=1}^N{\hat T}_p^p\right)^2+\sum_{p=1}^N(N-2p){\hat T}_p^p \ . 
\end{eqnarray}
\end{subequations}

As a possible boson realization of the above two algebras, 
two of the authors (J. da P. and M. Y.) and Kuriyama presented the form 
shown in Ref. \citen{12}: 
\begin{eqnarray}
& &{\hat S}^i={\hat a}_i^*{\hat b}+\sum_{p=1}^N{\hat a}^{p*}{\hat b}_i^p \ , 
\qquad
{\hat S}_i={\hat b}^*{\hat a}_i+\sum_{p=1}^N{\hat b}_i^{p*}{\hat a}^p \ , 
\nonumber\\
& &
{\hat S}_i^j={\hat a}_i^*{\hat a}_j-\sum_{p=1}^N{\hat b}_j^{p*}{\hat b}_i^p 
+\delta_{ij}\left(\sum_{p=1}^N {\hat a}^{p*}{\hat a}^p-{\hat b}^*{\hat b}
\right) \ , 
\label{2-3}\\
& &{\hat T}^p={\hat a}^{p*}{\hat b}^* 
-\sum_{i=1}^M{\hat a}_i^{*}{\hat b}_i^{p*} \ , 
\qquad
{\hat T}_p={\hat b}{\hat a}^p-\sum_{i=1}^M{\hat b}_i^{p}{\hat a}_i \ , 
\nonumber\\
& &
{\hat T}_q^p={\hat a}^{p*}{\hat a}^q 
+ \sum_{i=1}^M{\hat b}_i^{p*}{\hat b}_i^q 
+\delta_{pq}\left(\sum_{i=1}^M {\hat a}_i^{*}{\hat a}_i+{\hat b}^*{\hat b}
+(M+1)\right) \ . 
\label{2-4}
\end{eqnarray}
In the above forms, $({\hat a}_i , {\hat a}_i^*)$, 
$({\hat a}^p, {\hat a}^{p*})$, $({\hat b}_i^p, {\hat b}_i^{p*})$ and 
$({\hat b},{\hat b}^*)$ denote boson operators, where $i=1,2,\cdots , M$ 
and $p=1,2,\cdots , N$. 
We can prove that the forms (\ref{2-3}) and (\ref{2-4}) satisfy the 
relations (\ref{2-1}) and (\ref{2-2}), respectively. 
All the generators are expressed in terms of the bilinear forms for the boson 
operators. 
In relation to the $su(2)$-algebra in the original form,\cite{13} 
we call these forms the Schwinger boson representation. 
it should be noted that the two sets $({\hat S}^i,{\hat S}_i,{\hat S}_i^j)$ 
and $({\hat T}^p,{\hat T}_p,{\hat T}_q^p)$ commute mutually for 
any combination: 
\begin{equation}\label{2-5}
[\ {\hat S}^i \ , \ {\hat T}^p \ ]=0\ , \qquad {\rm etc.}
\end{equation}
Further, it should be noted that there exists an operator ${\hat R}$ which 
commutes with any operator defined in the forms (\ref{2-3}) and (\ref{2-4}):
\begin{equation}\label{2-6}
{\hat R}=\left(\sum_{p=1}^N {\hat a}^{p*}{\hat a}^p-{\hat b}^*{\hat b}\right)
-\sum_{i=1}^M\left({\hat a}_i^*{\hat a}_i-\sum_{p=1}^N{\hat b}_i^{p*}
{\hat b}_i^p\right) \ . 
\end{equation}
The operator ${\hat R}$ is related to ${\hat S}$ and ${\hat T}$ introduced 
in Ref. \citen{2} in the form 
\begin{equation}\label{2-7}
{\hat S}=-{\hat R} \ , \qquad 
{\hat T}=-{\hat R}-MN+1\ . 
\end{equation}
In this paper, we treat the case $N=M-1$. 
The reason will be later clarified. 
In Ref.\citen{14}, we treated the case $N=M$. 
In the framework of the above form, we investigate a possible counterpart 
of the Lipkin model in the $su(M+1)$-algebra for many-fermion system presented 
in the next section.

\section{The Lipkin model in the $su(M+1)$-algebra and its intrinsic states}

In this section, we recapitulate the Lipkin model in the $su(M+1)$-algebra 
developed in Ref.\citen{14}. 
Of course, this section contains a viewpoint different from that in 
Ref.\citen{14}. 
We consider a shell model consisting of $(M+1)$ single-particle levels, 
each of which is specified by a quantum number $i$ ($i=0,1,2,\cdots ,M)$. 
The degeneracy of each level is $\Omega$-fold. 
Then, introducing quantum number $\mu$ $(\mu=-\lambda, -\lambda+1, \cdots , 
\lambda-1, \lambda \ ; \ \lambda=$ half-integer, $\Omega=2\lambda+1)$, 
the single-particle state can be specified by ($i,\mu)$. 
In the above shell model for many-fermion system, we introduce the 
hole-operator $({\tilde \beta}_\mu , {\tilde \beta}_\mu^*)$ for the level 
$i=0$ and the particle-operator 
$({\tilde \alpha}_{i\mu},{\tilde \alpha}_{i\mu}^*)$ 
for the levels $i=1,2,\cdots , M$. 
With the use of the above fermion operators, the following operators can be 
defined: 
\begin{eqnarray}\label{3-1}
& &{\wtilde S}^i=\sum_\mu(-)^{\lambda-\mu}{\tilde \alpha}_{i\mu}^*
{\tilde \beta}_{-\mu}^* \ , \qquad
{\wtilde S}_i=\sum_\mu(-)^{\lambda-\mu}{\tilde \beta}_{-\mu}
{\tilde \alpha}_{i\mu} \ , \nonumber\\
& &{\wtilde S}_i^j=\sum_\mu {\tilde \alpha}_{i\mu}^*{\tilde \alpha}_{j\mu}
-\delta_{ij}(\Omega-\sum_{\mu}{\tilde \beta}_{\mu}^*{\tilde \beta}_\mu) \ .
\end{eqnarray}
We can prove that the form (\ref{3-1}) satisfies the relation (\ref{2-1}), 
i.e., a kind of the $su(M+1)$-algebraic models. 
If the single-particle level $i=0$ is fully occupied by fermions, the 
operator ${\wtilde S}^i$ describes particle-hole pair excitation.

First, we generalize the free vacuum in \S 1. 
For this purpose, let us presuppose that, in the present fermion space, 
there exist a state $\kket{m}$ uniquely, which is governed by the 
following condition: 
\begin{eqnarray}\label{3-2}
& &{\wtilde S}_i\kket{m}=0 \ , \qquad \hbox{\rm for\ any}\ i \ , \nonumber\\
& &{\wtilde S}_i^j\kket{m}=0 \ , \qquad \hbox{\rm for\ any}\ i\ {\rm and}\ j\ 
(j>i)\ . 
\end{eqnarray}
We call $\kket{m}$ the intrinsic state, which includes the free vacuum 
discussed in \S 1. 
In the terminology analogous to that appearing in nuclear rotational model, 
we use the intrinsic state. 
The excited states are obtained by operating the excited state generating 
operators on $\kket{m}$ appropriately. 
The detail will be discussed later. 
Operating with ${\wtilde S}_i^i$ on the both sides of the condition 
(\ref{3-2}) and using the commutation relation (\ref{2-1b}), we have 
\begin{eqnarray}\label{3-3}
& &{\wtilde S}_k\cdot{\wtilde S}_i^i \kket{m}=0 \ , 
\qquad \hbox{\rm for\ any}\ k\ {\rm and}\ i \ , \nonumber\\
& &{\wtilde S}_k^l\cdot{\wtilde S}_i^i\kket{m}=0 \ , 
\qquad \hbox{\rm for\ any}\ k, \ l\ {\rm and}\ i\ 
(l>k)\ . 
\end{eqnarray}
Under the relation (\ref{3-3}), the presupposition of the existence of 
the unique $\kket{m}$ leads to 
\begin{equation}\label{3-4}
{\wtilde S}_i^i\kket{m}=-\sigma_i\kket{m} \ . \qquad (i=1,2,\cdots , M)
\end{equation}
Here, $\sigma_i$ denotes $c$-number. 
The relation (\ref{3-4}) is nothing but the eigenvalue equation for the 
operator ${\wtilde S}_i^i$, and of course, $\sigma_i$ is real. 
The eigenvalue $\sigma_i$ is governed by the relation 
\begin{equation}\label{3-5}
0\leq \sigma_1 \leq \sigma_2 \leq \cdots \leq \sigma_M \ . 
\end{equation}
The condition (\ref{3-5}) is derived by the relation 
\begin{eqnarray}\label{3-6}
& &\bbra{m}{\wtilde S}_i\cdot{\wtilde S}^i\kket{m} \geq 0 \qquad
{\rm for\ any}\ i \ , \nonumber\\
& &\bbra{m}{\wtilde S}_i^j\cdot{\wtilde S}_j^i\kket{m} \geq 0 \qquad
{\rm for\ any}\ i \ {\rm and}\ j\ (j>i) \ . 
\end{eqnarray}
It may be self-evident that the state $\kket{m}$ is specified at least 
by $M$ quantum numbers and it is the eigenvalue of the Casimir operator: 
\begin{equation}\label{3-7}
\kket{m}=\kket{(\gamma);\sigma_1 \cdots \sigma_M} \ . 
\end{equation}
Here, $(\gamma)$ denotes a set of the quantum numbers which do not relate 
directly to the $su(M+1)$-algebra. 

In the set $(\gamma)$, total fermion number plays a special role. 
The operator of the total fermion number ${\wtilde N}$ is written down as 
\begin{equation}\label{3-8}
{\wtilde N}=\sum_{i=1}^M\left(\sum_\mu {\tilde \alpha}_{i\mu}^*
{\tilde \alpha}_{i\mu}\right)+
\left(\Omega-\sum_\mu{\tilde \beta}_{\mu}^*{\tilde \beta}_\mu\right) \ . 
\end{equation}
The operator ${\wtilde N}$ obeys 
\begin{equation}\label{3-9}
[\ {\wtilde N}\ , \ {\wtilde S}^i \ ]=[\ {\wtilde N}\ , \ {\wtilde S}_i \ ]
=[\ {\wtilde N}\ , \ {\wtilde S}_i^j \ ]=0 \ . 
\end{equation}
Further, it is important to see that ${\wtilde N}$ cannot be expressed in 
terms of a function of $({\wtilde S}^i,{\wtilde S}_i,{\wtilde S}_i^j)$. 
Therefore, in addition to the eigenvalue equation for ${\wtilde S}_i^i$, 
independently, we can set up the eigenvalue equation: 
\begin{equation}\label{3-10}
{\wtilde N}\kket{m}={\cal N}\kket{m} \ . 
\end{equation}
Therefore, $\kket{m}$ is regarded as a state with ${\cal N}$ fermions. 
The operators $(\Omega-\sum_\mu{\tilde \beta}_\mu^*{\tilde \beta}_\mu)$ 
and 
$\sum_\mu{\tilde \alpha}_{i\mu}^*{\tilde \alpha}_{i\mu}$ are expressed as 
\begin{eqnarray}\label{3-11}
& &\Omega-\sum_\mu{\tilde \beta}_\mu^*{\tilde \beta}_\mu
=(M+1)^{-1}\left({\wtilde N}-\sum_{j=1}^M{\wtilde S}_j^j\right) \ , \nonumber\\
& &\sum_\mu{\tilde \alpha}_{i\mu}^*{\tilde \alpha}_{i\mu}=
(M+1)^{-1}\left({\wtilde N}-\sum_{j=1}^M{\wtilde S}_j^j\right)
+{\wtilde S}_i^i \ . 
\end{eqnarray}
Then, the following eigenvalue equation is derived: 
\begin{subequations}\label{3-12}
\begin{eqnarray}
& &\left(\Omega-\sum_\mu{\tilde \beta}_\mu^*{\tilde \beta}_\mu\right)
\kket{m}=(\Omega-n)\kket{m} \ , \nonumber\\
& &\left(\sum_\mu{\tilde \alpha}_{i\mu}^*{\tilde \alpha}_{i\mu}\right)
\kket{m}=n_i\kket{m} \ , 
\label{3-12a}\\
& &\Omega-n=(M+1)^{-1}\left({\cal N}+\sum_{j=1}^M \sigma_j\right) \ , 
\nonumber\\
& &n_i=(M+1)^{-1}\left({\cal N}+\sum_{j=1}^M \sigma_j\right)-\sigma_i \ . 
\label{3-12b}
\end{eqnarray}
Inversely, we have 
\begin{equation}\label{3-12c}
\sigma_i=\Omega-n-n_i \ , \qquad {\cal N}=\Omega-n+\sum_{i=1}^M n_i \ . 
\end{equation}
\end{subequations}

As is clear from the above argument, the total fermion number ${\cal N}$ 
is essential for understanding fermion number distribution in the 
intrinsic state $\kket{m}$ of the Lipkin model in the $su(M+1)$-algebra, 
but, it is a quantum number additional to those which are proper to the 
$su(M+1)$-algebra, i.e., $(\sigma_1,\sigma_2,\cdots ,\sigma_M)$. 
Therefore, it can be concluded that the minimum number of the 
quantum numbers specifying $\kket{m}$ for the $su(M+1)$-algebra is $M$. 
Concerning the concrete expression of $\kket{m}$, we can find it easily 
for the case of the free vacuum: 
\begin{equation}\label{3-13}
\kket{m}=\prod_{i=1}^l \left(\prod_{\mu=-\lambda}^\lambda 
{\tilde \alpha}_{i\mu}^*\right)\kket{0} \ . \qquad
({\tilde \alpha}_{i\mu}\kket{0}={\tilde \beta}_\mu\kket{0}=0)
\end{equation}
The state (\ref{3-13}) satisfies the condition (\ref{3-2}) and 
we have 
\begin{equation}\label{3-14}
n=0\ , \quad n_i=\Omega \quad {\rm for}\ i=1,\cdots , l \ , \qquad
n_i=0 \quad {\rm for}\ i=l+1,\cdots ,M \ . 
\end{equation}
In the present case, the excited state generating operators which we denote 
as ${\wtilde \Phi}^i$ and ${\wtilde \Phi}_j^i$ may be given as 
\begin{eqnarray}\label{3-15}
& &{\wtilde \Phi}^i={\wtilde S}^i \ , \qquad (l+1 \leq i \leq M) \nonumber\\
& &{\wtilde \Phi}_j^i={\wtilde S}_j^i \ . \qquad 
(1 \leq i \leq l \ , \ l+1\leq j \leq M)
\end{eqnarray}
The form (\ref{3-15}) is suggested by the particle-hole pair Tamm-Dankoff 
approximation. 
However, practically, it may be almost impossible to obtain the explicit 
form of $\kket{m}$ except for the form (\ref{3-13}), because we must specify 
the set of the quantum numbers $(\gamma)$ except for ${\cal N}$ for 
$\kket{m}$, and further, it may be almost impossible to specify the 
excited state generating operators. 
The above consideration leads to the following viewpoint: 
The Lipkin model in the $su(M+1)$-algebra is workable for the description of 
collective motion and its coupling to non-collective degrees of freedom 
based on the free vacuum. 
However, for the study of finite temperature effects, this model does not 
work in the framework shown in this section, because all the intrinsic 
states are necessary. 
Later, we will contact this problem.

\section{The Holstein-Primakoff boson representation for the $su(M+1)$-algebra}

As a possible boson realization of the $su(M+1)$-algebra, we know the 
Holstein-Primakoff boson representation, the original form of which is 
for the $su(2)$-algebra.\cite{15} 
In this representation, the generators $({\ol S}^i,{\ol S}_i,{\ol S}_i^j)$ can 
be expressed as follows:
\begin{eqnarray}\label{4-1}
& &{\ol S}^i={\hat c}_i^*\sqrt{\sigma-\sum_{k=1}^M{\hat c}_k^*{\hat c}_k} \ , 
\qquad
{\ol S}_i=\sqrt{\sigma-\sum_{k=1}^M{\hat c}_k^*{\hat c}_k}\cdot {\hat c}_i \ , 
\nonumber\\
& &{\ol S}_i^j={\hat c}_i^*{\hat c}_j
-\delta_{ij}\left(\sigma-\sum_{k=1}^M{\hat c}_k^*{\hat c}_k\right) \ . 
\end{eqnarray}
Here, $\sigma$ denotes a positive integer. 
The operator $({\hat c}_i, {\hat c}_i^*)$ denotes the boson operator. 
Of course, the form (\ref{4-1}) obeys the $su(M+1)$-algebra. 

An intrinsic state $\rket{m}$ in this representation is the vacuum for 
${\hat c}_i$: 
\begin{equation}\label{4-2}
\rket{m}=\rket{0} \ . \qquad ({\hat c}_i\rket{0}=0)
\end{equation}
The state $\rket{m}$ obeys 
\begin{eqnarray}
& &{\ol S}_i\rket{m}=0 \quad {\rm for\ any}\ i \ , \nonumber\\
& &{\ol S}_i^j\rket{m}=0 \quad {\rm for\ any}\ i\ , j \ (j>i) \ , 
\label{4-3}\\
& &{\ol S}_i^i\rket{m}=-\sigma_i\rket{m} \ , \quad \sigma_i=\sigma \ . 
\label{4-4}
\end{eqnarray}

From the above treatment, it may be clear that there does not exist any 
possibility to introduce new quantum number such as ${\cal N}$. 
Then, in order to regard the present representation as a counterpart 
of the Lipkin model in the $su(M+1)$-algebra, we introduce the quantity 
$(\Omega-n)$ as a parameter. 
Since any of $\sigma_i$ is equal to $\sigma$, we can define $n_i$ as follows: 
\begin{equation}\label{4-5}
n_i=\Omega-n-\sigma \ . \qquad (i=1,2,\cdots ,M)
\end{equation}
Therefore, ${\cal N}$ is given in the form 
\begin{equation}\label{4-6}
{\cal N}=(M+1)(\Omega-n)-M\sigma \ . 
\end{equation}
We can see that $n_i$ does not depend on $i$ and if $n=n_i=0$, 
we obtain 
\begin{equation}\label{4-7}
\sigma={\cal N}=\Omega \ . 
\end{equation}
The above means that the single-particle level $i=0$ is occupied fully 
by $\Omega$ fermions and all the other levels are vacant, i.e., 
a closed shell system. 
It is well known fact and usually this case has been investigated. 
However, it is noted that the applicability of the Holstein-Primakoff 
boson representation is wider than the case (\ref{4-7}). 
We can treat the following case: $(\Omega-n)$ fermions occupy the 
single-particle level $i=0$ and all the others are occupied by 
$(\Omega-n-\sigma)$ fermions for each level. 
However, it should be noted that, in the framework of the Holstein-Primakoff 
boson representation, it is impossible to treat the case such as 
the case where two levels $i=0$ and $1$ are occupied and the others are 
vacant. 
The aim of this paper is to present an idea, with the aid of which the 
above cases can be treated.

In the next section, we will investigate the above-mentioned problem and 
as a preliminary argument, we will show the same results as the above 
in the framework of the Schwinger boson representation presented in \S 2. 
As the intrinsic state $\ket{m}$, we can choose the following form: 
\begin{equation}\label{4-8}
\ket{m}=({\hat b}^*)^{\sigma}\ket{0} \ . 
\end{equation}
Here, $\ket{0}$ denotes the vacuum for ${\hat b}$, ${\hat b}_i^p$ and 
${\hat a}^p$. 
The state $\ket{m}$ satisfies the same relations as those shown in 
Eqs.(\ref{4-3}) and (\ref{4-4}). 
This case also has no possibility to introduce new quantum numbers. 
The state (\ref{4-2}) or (\ref{4-8}) corresponds to the symmetric 
representation for the $su(M+1)$-algebra, and then, we have to contact with 
the case of non-symmetric representation.

\section{The intrinsic state in the Schwinger boson representation}

In the boson space constructed in terms of the bosons $({\hat a}_i^*,
{\hat a}^{p*},{\hat b}_i^{p*},{\hat b}^*\ ; \ i=1,2,\cdots ,M , \ p=1,2,\cdots 
,N)$, we presuppose that there exists a state $\ket{m}$ uniquely, which 
is governed by the following condition: 
\begin{subequations}\label{5-1}
\begin{eqnarray}
& &{\hat S}_i\ket{m}=0 \quad {\rm for\ any}\ i\ , \qquad
{\hat S}_i^j\ket{m}=0 \quad {\rm for\ any}\ i\ {\rm and}\ j\ (j>i) \ , 
\label{5-1a}\\
& &{\hat T}_p\ket{m}=0 \quad {\rm for\ any}\ p\ , \qquad
{\hat T}_q^p\ket{m}=0 \quad {\rm for\ any}\ p\ {\rm and}\ q\ (q>p) \ . 
\label{5-1b}
\end{eqnarray}
\end{subequations}
The condition (\ref{5-1a}) corresponds to the condition (\ref{3-2}). 
If $\ket{m}$ is obtained in $(M+m)$ quantum numbers, we can construct the 
excited states in the present boson space by operating 
${\hat S}^i$, ${\hat S}_j^i$ ($j>i$), ${\hat T}^p$ and ${\hat T}_p^q$ 
($q>p)$ appropriately on the state $\ket{m}$. 
If this idea is acceptable, the total number of the quantum numbers is 
given by $(M^2+M)/2+(N^2+N)/2+(M+m)$. 
On the other hand, the present boson space consists of 
$(M+1)(N+1)$ kinds of boson operators. 
Then, we have 
\begin{equation}\label{5-2}
(M^2+M)/2+(N^2+N)/2+(M+m)=(M+1)(N+1) \ . 
\end{equation}
The relation (\ref{5-2}) gives us 
\begin{equation}\label{5-3}
N=M+(1\pm\sqrt{9-8m})/2 \ . 
\end{equation}
The above leads to 
\begin{eqnarray}
& &N=M,\ M+1 \quad {\rm for}\ m=1 \ , 
\label{5-4}\\
& &N=M-1, \ M+2 \quad {\rm for}\ m=0 \ . 
\label{5-5}
\end{eqnarray}
The case $N=M$ was treated in Ref.\citen{14}. 
In this case, $\ket{m}$ consists of $(M+1)$ quantum numbers, 
in which the $M$ quantum numbers are related to the eigenvalues of 
${\hat S}_i^i$ ($i=1,2,\cdots , M$) and the last one is related to the 
eigenvalue of a certain operator expected to play the same role as that 
of ${\wtilde N}$ in the Lipkin model in the $su(M+1)$-algebra. 
The form obtained in this framework is too complicated to come up 
to our expectation. 
Then, in this paper, we treat the case 
\begin{equation}\label{5-6}
N=M-1 \ . 
\end{equation}
In this case, $\ket{m}$ is specified by $M$ quantum numbers. 
Then, in order to obtain a counterpart of the Lipkin model in the 
$su(M+1)$-algebra, as a parameter, we introduce the quantity 
$(\Omega-n)$ in the same way as that in \S 4.

Let us construct the state $\ket{m}$. 
First, we note that the condition (\ref{5-1}) suggests us the following form 
for $\ket{m}$: 
$\ket{m}$ is expressed only in terms of the operators ${\hat b}^*$ and 
${\hat b}_i^{p*}$ ($i=1,2,\cdots , M ,\ p=1,2,\cdots ,M-1)$. 
Under this note, we introduce the operator ${\hat b}_k(M)^*$ 
($k=0,1,\cdots ,M-1)$ defined as 
\begin{subequations}\label{5-7}
\begin{eqnarray}
& &{\hat b}_0(M)^*={\hat b}^* \ , 
\label{5-7a}\\
& &{\hat b}_k(M)^*=\left|
\begin{array}{@{\,}cccc@{\,}}
{\hat b}_M^{1*} & {\hat b}_M^{2*} & \cdots & {\hat b}_M^{M-k *} \\
{\hat b}_{M-1}^{1*} & {\hat b}_{M-1}^{2*} & \cdots & {\hat b}_{M-1}^{M-k *} 
\\
\cdots & \cdots & \cdots & \cdots \\
{\hat b}_{k+1}^{1*} & {\hat b}_{k+1}^{2*} & \cdots & {\hat b}_{k+1}^{M-k *} \\
\end{array}
\right| \ . 
\quad (k=1,2,\cdots , M-1) 
\label{5-7b}
\end{eqnarray}
\end{subequations}
We can prove the following relation for any $(i,j\ ; \ j>i)$ and $k$: 
\begin{equation}\label{5-8}
[\ {\hat S}_i^j \ , \ {\hat b}_k(M)^* \ ]=
-\sum_p [\ {\hat b}_j^{p*}{\hat b}_i^{p} \ , \ {\hat b}_k(M)^* \ ]=0 \ .
\end{equation}
For the case ${\hat b}_0(M)^*$ ($={\hat b}^*)$, the relation (\ref{5-8}) 
is trivial. 
In the case $i\leq k$, ${\hat b}_k(M)^*$ does not contain ${\hat b}_i^{p*}$ 
and we have the relation (\ref{5-8}). 
In the case $k+1 \leq i < M$, any component of the $i$-the row of 
the determinant (\ref{5-8}) is replaced with the component in the 
$j$-th row, and then, the determinant (\ref{5-8}) vanishes. 
The case $i=M$ does not exist because $i<j$ and $j \leq M$. 
In the same way as in the above, we can prove 
\begin{equation}\label{5-9}
[\ {\hat T}_q^{p} \ , \ {\hat b}_k(M)^*\ ]=0\quad {\rm for\ any}\ 
p,\ q\ {\rm and}\ k\ (q>p). 
\end{equation}
With the use of ${\hat b}_k(M)^*$, we define $\ket{m}$ in the form 
\begin{equation}\label{5-10}
\ket{m}=\ket{m_0,m_1,\cdots , m_{M-1}}
=\prod_{k=0}^{M-1}({\hat b}_k(M)^*)^{m_k}\ket{0} \ . 
\end{equation}
Here, the normalization constant is omitted. 
Since ${\hat S}_i$ and ${\hat T}_p$ contain the annihilation operators 
${\hat a}_i$ and ${\hat a}^p$ and the state $\ket{m}$ is constructed 
in terms of ${\hat b}^*$ and ${\hat b}_i^{p*}$, we have the first of the 
relation (\ref{5-1}). 
With the use of the relations (\ref{5-8}) and (\ref{5-9}), 
we also have the second of the relation (\ref{5-1}). 
Therefore, the state (\ref{5-10}) can be regarded as $\ket{m}$. 

We can prove that the state (\ref{5-10}) is an eigenstate of the set 
$({\hat S}_i^i\ ; \ i=1,2,\cdots ,M)$: 
\begin{eqnarray}
& &{\hat S}_i^i\ket{m_0,m_1,\cdots ,m_{M-1}}
=-\sigma_i\ket{m_0,m_1,\cdots ,m_{M-1}} \ , 
\label{5-11}\\
& &\sigma_i=\sum_{k=0}^{i-1}m_k \ . 
\label{5-12}
\end{eqnarray}
For the proof of the relation (\ref{5-11}) with the eigenvalue (\ref{5-12}), 
the following relation is available:
\begin{eqnarray}\label{5-13}
& &[\ {\hat S}_i^i \ , \ {\hat b}_0(M)^* \ ]=-{\hat b}_0(M)^* \ , \nonumber\\
& &[\ {\hat S}_i^i \ , \ {\hat b}_k(M)^* \ ]=0 \ , \quad (1\leq i \leq k) 
\nonumber\\
& &[\ {\hat S}_i^i \ , \ {\hat b}_k(M)^* \ ]=-{\hat b}_k(M)^* \ . 
\quad (k+1 \leq i \leq M)
\end{eqnarray}
We, further, prove the following relation: 
\begin{eqnarray}
& &{\hat T}_p^p\ket{m_0,m_1,\cdots ,m_{M-1}}
=\tau_p\ket{m_0,m_1,\cdots ,m_{M-1}} \ , 
\label{5-14}\\
& &\tau_p=\sum_{k=0}^{M-p}m_k +(M+1) \ . 
\label{5-15}
\end{eqnarray}
In the same way as in the case of ${\hat S}_i^i$, the following relation is 
useful for the proof:
\begin{eqnarray}\label{5-16}
& &[\ {\hat T}_p^p \ , \ {\hat b}_0(M)^* \ ]={\hat b}_0(M)^* \ , \nonumber\\
& &[\ {\hat T}_p^p \ , \ {\hat b}_k(M)^* \ ]={\hat b}_k(M)^* \ , 
\quad (1\leq p \leq M-k) 
\nonumber\\
& &[\ {\hat T}_p^p \ , \ {\hat b}_k(M)^* \ ]=0 \ . 
\quad (M-k+1 \leq p \leq M-1)
\end{eqnarray}
The proof of the relations (\ref{5-13}) and (\ref{5-16}) is straightforward. 
Thus, we are able to obtain the state $\ket{m}$ under the condition 
(\ref{5-1}), and further, in $M$ quantum numbers.

\section{Properties of the intrinsic state in the Schwinger boson 
representation}

First, we note the relations (\ref{5-12}) and (\ref{5-15}). 
These relations give 
\begin{eqnarray}
& &\tau_p=\sigma_{M-p+1}+(M+1) \ , 
\label{6-1}\\
& &m_0=\sigma_1\ , \qquad m_k=\sigma_{k+1}-\sigma_k \ . \quad
(k=1,2,\cdots ,M-1) 
\label{6-2}
\end{eqnarray}
Therefore, the state $\ket{m}$ is specified in terms of the eigenvalue 
of ${\hat S}_i^i$ ($i=1,2,\cdots ,M)$. 
It may be clear that there does not exist any possibility to add one more 
quantum number and $\ket{m}$ is also the eigenstate of ${\hat T}_p^p$ 
($p=1,2,\cdots,M-1)$. 
Therefore, in order to connect the Schwinger boson representation in the 
$su(M+1)$-algebra shown in \S 5 with the Lipkin model in the 
$su(M+1)$-algebra, we must introduce a quantity $(\Omega-n)$ as a 
parameter from the outside and define $n_i$ and ${\cal N}$ in the 
following forms:
\begin{eqnarray}
& &n_i=\Omega-n-\sigma_i=\Omega-n-\sum_{k=0}^{i-1}m_k \ , 
\label{6-3}\\
& &{\cal N}=\Omega-n+\sum_{i=1}^M n_i=(M+1)(\Omega-n)
-\sum_{k=0}^{M-1}(M-k)m_k\ . 
\label{6-4}
\end{eqnarray}

\begin{figure}[t]
  \epsfxsize=8cm  
  \centerline{\epsfbox{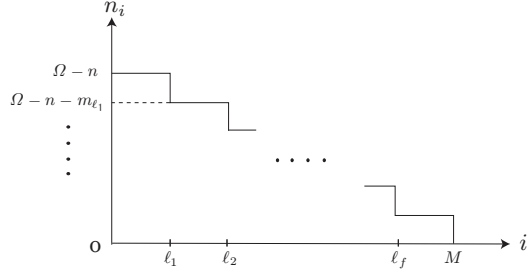}}
  \caption{The relation (\ref{6-8}) for $n_i$ is schematically shown.}
   \label{fig1}
\end{figure}

Let us, first, examine the relation (\ref{6-3}) in the case 
\begin{equation}\label{6-5}
n=0\ , \qquad m_l=\Omega\ , \qquad m_k=0 \ \  (k\neq l) \ .
\end{equation}
We can see that the relation (\ref{6-5}) is reduced to the relation 
(\ref{3-14}). 
Therefore, the following state is a counterpart of $\kket{m}$ shown in the 
relation (\ref{3-13}):
\begin{equation}\label{6-6}
\ket{m}=({\hat b}_l(M)^*)^\Omega \ket{0} \ . 
\end{equation}
The state (\ref{6-6}) plays a role of the free vacuum in which the 
single-particle levels $i=0,1,2,\cdots ,l$ are fully occupied by 
fermions and the levels $i=l+1,l+2,\cdots M$ vacant. 
Next, let us investigate a case which is much more complicated than the 
previous one: 
\begin{equation}\label{6-7}
m_k=0 \quad {\rm for}\quad k\neq l_1,l_2,\cdots , l_f \ . \ \ 
(l_1<l_2<\cdots <l_f)
\end{equation}
The relation (\ref{6-3}) under the case (\ref{6-7}) gives us 
\begin{eqnarray}\label{6-8}
& &n_1=n_2= \quad \cdots \quad=n_{l_1}=\Omega-n \ , \nonumber\\
& &n_{l_1+1}=n_{l_1+2}=\cdots =n_{l_2}=\Omega-n-m_{l_1} \ , \nonumber\\
& &n_{l_2+1}=n_{l_2+2}=\cdots =n_{l_3}=\Omega-n-(m_{l_1}+m_{l_2}) 
\ , \nonumber\\
& & \qquad\qquad\qquad\qquad \vdots \nonumber\\
& &n_{l_f+1}=n_{l_f+2}=\cdots =n_{M}=\Omega-n-(m_{l_1}+\cdots +m_{l_f}) \ . 
\end{eqnarray}
The relation (\ref{6-8}) can be shown schematically in Fig.\ref{fig1}. 
%
%
Figure 1 suggests that we can describe various cases of fermion number 
distributions, for example, non-closed shell state including the case 
in which the fermion number is equal to $l\Omega$ ($l$: integer), but, the 
fermion number distribution is not of the closed shell form. 
Of course, the help of the non-symmetric representation is necessary.

It may be an interesting problem to investigate the excited state generating 
operators which construct the excited states from the intrinsic state 
$\ket{m}$. 
The operators ${\hat S}_i$ and ${\hat S}_i^j$ ($j>i$) are used to determine 
the state $\ket{m}$. 
This means that for obtaining the excited states, it may be necessary to 
investigate ${\hat S}^i\ (={\hat S}_i^*$) and ${\hat S}_j^i\ (=
{\hat S}_i^{j*};\ j>i$). 
Since we are interested in the $su(M+1)$-algebra, we will not contact 
with ${\hat T}^p$ and ${\hat T}_p^q\ (q>p)$ in this paper. 
For this problem, the following relation should be noted: 
\begin{subequations}\label{6-9}
\begin{eqnarray}
& &[\ {\hat S}^i \ , \ {\hat b}_k(M)^*\ ]=0 \ , \quad (1\leq i\leq k) 
\label{6-9a}\\
& &[\ {\hat S}_j^i \ , \ {\hat b}_k(M)^*\ ]=0 \ . \quad (1\leq i <j \leq k \ , 
\ k+1 \leq i<j<M) 
\label{6-9b}
\end{eqnarray}
\end{subequations}
The proof is in parallel to the case of the relation (\ref{5-8}). 
In order to reconfirm the relation (\ref{3-15}) for the Lipkin model 
in the $su(M+1)$-algebra, which is suggested by the particle-hole pair 
Tamm-Dankoff approximation, we show the case (\ref{6-5}), i.e., (\ref{3-14}). 
The general formula (\ref{6-9}) suggests for the case (\ref{6-5}) to divide 
${\hat S}^i$ and ${\hat S}_j^i\ (j>i)$ into two groups: 
\begin{subequations}\label{6-10}
\begin{eqnarray}
& &{\hat S}^i=\biggl\{
\begin{array}{@{\,}ll@{\,}}
{\hat \Psi}^i & (1\leq i \leq l) \\ 
{\hat \Phi}^i & (l+1\leq i \leq M)
\end{array} \ , 
\label{6-10a}\\
& &{\hat S}_j^i=\biggl\{
\begin{array}{@{\,}ll@{\,}}
{\hat \Psi}_j^i &  (1\leq i<j \leq l \ , \ l+1\leq i<j\leq M) \\
{\hat \Phi}_j^i &  (1\leq i \leq l, \ l+1\leq j \leq M)
\end{array} \ . 
\label{6-10b}
\end{eqnarray}
\end{subequations}
For the state (\ref{6-6}), the relation (\ref{6-9}) gives 
\begin{subequations}\label{6-11}
\begin{eqnarray}
& &{\hat \Psi}^i\ket{m}={\hat \Psi}_j^i \ket{m}=0\ , 
\label{6-11a}\\
& &{\hat \Phi}^i\ket{m} \neq 0 \ , \qquad {\hat \Phi}_j^i\ket{m} \neq 0 \ . 
\label{6-11b}
\end{eqnarray}
\end{subequations}
Further, we have 
\begin{subequations}\label{6-12}
\begin{eqnarray}
& &[\ {\hat \Psi}^{\alpha} \ , \ {\hat \Phi}^i \ ]=0\ , \qquad
[\ {\hat \Psi}^{\alpha} \ , \ {\hat \Phi}_j^i \ ]
=-\delta_{\alpha i}{\hat \Phi}^j\ , \nonumber\\
& &[\ {\hat \Psi}_{\alpha}^{\beta} \ , \ {\hat \Phi}^i \ ]
=\delta_{\beta i}{\hat \Psi}^{\alpha}\ , \qquad
[\ {\hat \Psi}_{\alpha}^{\beta} \ , \ {\hat \Phi}_j^i \ ]=0\ , 
\label{6-12a}\\
& &[\ {\hat \Phi}^{\alpha} \ , \ {\hat \Phi}^i \ ]=
[\ {\hat \Phi}_{\alpha}^{\beta} \ , \ {\hat \Phi}^i \ ]=
[\ {\hat \Phi}_{\alpha}^{\beta} \ , \ {\hat \Phi}_j^i \ ]=0\ . 
\label{6-12b}
\end{eqnarray}
\end{subequations}
The relations (\ref{6-11}) and (\ref{6-12}) tell us that 
${\hat \Phi}^i$ and ${\hat \Phi}_j^i$ play a role of the excited state 
generating operators from $\ket{m}$ shown in the relation (\ref{6-6}). 
This supports the relation (\ref{3-15}). 
However, in this paper, we do not contact with the problem how to 
construct orthogonalized excited states. 
Total number of ${\hat \Phi}^i$ and ${\hat \Phi}_j^i$ is 
$(M-l)(l+1)$. 
If $m_0 \neq 0$, $m_l \neq 0$ and $m_k=0$ ($k\neq 0,l$), the excited state 
generating operators are ${\hat \Phi}^i$ ($1\leq i \leq M$) and 
${\hat \Phi}_j^i$ ($1\leq i \leq l,\ l+1\leq j \leq M$). 
In this case, total number of ${\hat \Phi}^i$ and ${\hat \Phi}_j^i$ 
is $(M+l(M-l))$, and then, we can see that depending on the structure 
of $\ket{m}$, the number of ${\hat \Phi}^i$ and ${\hat \Phi}_j^i$ 
becomes different. 
The extension to more general cases may be straightforward.

As was already mentioned, the intrinsic state $\ket{m}$ is specified 
completely by the set of the quantum numbers 
($\sigma_1, \sigma_2, \cdots ,\sigma_M)$.
On the other hand, the intrinsic state $\kket{m}$ is specified 
not only by ($\sigma_1, \sigma_2, \cdots ,\sigma_M)$ but also by 
$(\gamma)$. 
Clearly, both are not in the relation of one to one correspondence. 
Therefore, if $\ket{m}$ is used to describe collective motion and 
its coupling to non-collective degrees of freedom, we do not have any obstacle. 
In this case, the description is based on a chosen intrinsic state 
from all $\ket{m}$. 
The above offers us a question if $\ket{m}$ is useful for describing 
the finite temperature effects or not. 
The reply is as follows: 
If any physical quantity in which we are interested is expressed as a 
function of the $su(M+1)$-generators, the intrinsic state $\ket{m}$ is 
useful. 
The outline of the reason is given in Appendix.

\section{Discussion}

Finally, we discuss the simplest three cases, i.e., $(M=1, N=0)$, 
$(M=2, N=1)$ and ($M=3, N=2)$. 
The case $(M=1, N=0)$ corresponds to the $su(2)$-algebra and the 
generators are expressed as 
\begin{equation}\label{7-1}
{\hat S}^1={\hat a}_1^*{\hat b}\ , \qquad
{\hat S}_1={\hat b}^*{\hat a}_1 \ , \qquad
{\hat S}_1^1={\hat a}_1^*{\hat a}_1-{\hat b}^*{\hat b}\ . 
\end{equation}
Under the change of the notations ${\hat S}^1={\hat S}_+$, 
${\hat S}_1={\hat S}_-$ and 
$(1/2){\hat S}_1^1={\hat S}_0$, we can see that $({\hat S}_{\pm,0}$) forms the 
$su(2)$-algebra in the original Schwinger boson representation.\cite{13} 
The intrinsic state is only of the form 
\begin{equation}\label{7-2}
\ket{m}=({\hat b}_0(1)^*)^{m_0}\ket{0}
=({\hat b}^*)^{m_0}\ket{0} \ . \quad 
(m_0=0,1,\cdots)
\end{equation}
Further, the excited state generating operator is ${\hat \Phi}^1={\hat S}^1
={\hat S}_+$. 
Then, under the change of the notations ${\hat a}_1={\hat a}$ and 
$m_0=2s$, the orthogonal set is given in the conventional notations by 
\begin{eqnarray}\label{7-3}
& &\ket{s}=({\hat b}^*)^{2s}\ket{0} \ , \nonumber\\
& &\ket{s,s_0}=({\hat S}_+)^{s+s_0}\ket{s}
=({\hat a}^*)^{s+s_0}({\hat b}^*)^{s-s_0} \ket{0} \ , \nonumber\\
& &\qquad\qquad (s=0,1/2,1,\cdots \ ; \ s_0=-s, -s+1,\cdots ,s-1,s)
\end{eqnarray}
The relations (\ref{6-3}) and (\ref{6-4}) lead us to 
\begin{equation}\label{7-4}
s=(\Omega-n)-{\cal N}/2\ , \qquad n_1={\cal N}-(\Omega-n)\ . 
\end{equation}
The relation (\ref{7-4}) shows 
\begin{equation}\label{7-5}
s=\Omega/2-n \ , \qquad n_1=n\quad {\rm for}\ {\cal N}=\Omega\ . 
\end{equation}
In relation to the applicability of the $su(2)$-algebra in the Schwinger boson 
representation to the finite temperature effects, in Appendix, we discuss 
the connection of the state $\ket{m}$ shown in the relation (\ref{7-2}) 
or (\ref{7-3}) to $\kket{m}$ in the Lipkin model in the $su(2)$-algebra 
in detail.

Next, we discuss the case $(M=2, N=1)$, i.e., the 
$su(3)$-algebra. 
The generators are obtained by putting $i,j=1,2$ and $p=1$ in the 
expression (\ref{2-1}). 
In this case, the intrinsic states are expressed in terms of two operators: 
\begin{equation}\label{7-6}
{\hat b}_0(2)^*={\hat b}^* \ , \qquad {\hat b}_1(2)^*={\hat b}_2^{1*} \ . 
\end{equation}
Further, the excited state generating operators are ${\hat \Phi}^1={\hat S}^1$ 
and ${\hat \Phi}^2={\hat S}^2$ for the state 
\begin{equation}\label{7-7}
\ket{m}=({\hat b}_0(2)^*)^{m_0}\ket{0}=({\hat b}^*)^{m_0}\ket{0} \ . 
\end{equation}
Also, ${\hat \Phi}_2^1={\hat S}_2^1$ and ${\hat \Phi}^2={\hat S}^2$ are the 
excited state generating operators for the case 
\begin{equation}\label{7-8}
\ket{m}=({\hat b}_1(2)^*)^{m_1}\ket{0}=({\hat b}_2^{1*})^{m_1}\ket{0} \ . 
\end{equation}
For both cases, number of the excited state generating operators is two. 
Let ${\hat a}^1$, ${\hat a}_1$, ${\hat a}_2$, ${\hat b}$, ${\hat b}_1^1$ and 
${\hat b}_2^1$ read 
\begin{equation}\label{7-9}
{\hat a}^1 \rightarrow {\hat a}_2 \ , \quad
{\hat a}_1 \rightarrow {\hat b}_1^1\ , \quad
{\hat a}_2 \rightarrow {\hat a}^1 \ , \quad
{\hat b}\rightarrow {\hat b}_2^1 \ , \quad
{\hat b}_1^1 \rightarrow -{\hat a}_1 \ , \quad 
{\hat b}_2^1\rightarrow {\hat b} \ .
\end{equation}
Then, ${\hat S}_2^1$ and ${\hat S}^2$ for the state (\ref{7-8}) becomes 
${\hat S}^1$ and ${\hat S}^2$ for the state (\ref{7-7}): 
\begin{equation}\label{7-10}
{\hat S}_2^1 \rightarrow {\hat S}^1\ , \qquad 
{\hat S}^2 \rightarrow {\hat S}^2 \ . 
\end{equation}
The above means that, by relabeling the operators, the case (\ref{7-8}) 
is reduced to the case (\ref{7-7}). 
Therefore, the case (\ref{7-8}) is also in the symmetric representation. 
In the case of the Lipkin model in the $su(M+1)$-algebra, the same 
situation is observed by performing the particle-hole conjugation. 
If $\ket{m}=({\hat b}_0(2)^*)^{m_0}\cdot ({\hat b}_1(2)^*)^{m_1}\ket{0}$, 
the excited state generating operators are ${\hat \Phi}^1={\hat S}^1$, 
${\hat \Phi}^2={\hat S}^2$ and ${\hat \Phi}_2^1={\hat S}_2^1$. 
The quantum numbers appearing in the Elliott's $su(3)$-model,\cite{16} 
$\lambda$ and $\mu$, are expressed as $\lambda=m_0+m_1/2$ and $\mu=m_1/2$. 
A further study of the $su(3)$-algebra will be reported in 
subsequent paper.

A main interest of this section is concerned with the case $(M=3, N=2)$, 
which corresponds to the $su(4)$-algebra. 
The intrinsic states are constructed by the following operators: 
\begin{eqnarray}
& &{\hat b}_0(3)^*={\hat b}^* \ , \qquad {\hat b}_2(3)^*={\hat b}_3^{1*} \ , 
\label{7-11}\\
& &{\hat b}_1(3)^*=
\left| 
\begin{array}{@{\,}cc@{\,}}
{\hat b}_3^{1*} & {\hat b}_3^{2*} \\
{\hat b}_2^{1*} & {\hat b}_2^{2*}
\end{array}
\right|
={\hat b}_3^{1*}{\hat b}_2^{2*}-{\hat b}_3^{2*}{\hat b}_2^{1*} \ . 
\label{7-12}
\end{eqnarray}
In this case, the excited state generating operators are 
${\hat \Phi}^1={\hat S}^1$, ${\hat \Phi}^2={\hat S}^2$ and 
${\hat \Phi}^3={\hat S}^3$ for the state 
\begin{subequations}\label{7-13}
\begin{equation}\label{7-13a}
\ket{m}=({\hat b}_0(3)^*)^{m_0}\ket{0}=({\hat b}^*)^{m_0}\ket{0} \ . 
\end{equation}
On the other hand, ${\hat \Phi}_3^1={\hat S}_3^1$, 
${\hat \Phi}_3^2={\hat S}_3^2$ and ${\hat \Phi}^3={\hat S}^3$ are the 
excited state generating operators in the case 
\begin{equation}\label{7-13b}
\ket{m}=({\hat b}_2(3)^*)^{m_2}\ket{0}=({\hat b}_3^{1*})^{m_2}\ket{0} \ . 
\end{equation}
\end{subequations}
In both cases, the number of the excited state generating operators is three 
and by relabeling the boson operators in a way similar to the case of 
$su(3)$-algebra, one is reduced to the other. 
Both are in the symmetric representation. 
The excited state generating operators are 
${\hat \Phi}^2={\hat S}^2$, ${\hat \Phi}^3={\hat S}^3$, 
${\hat \Phi}_2^1={\hat S}_2^1$ and ${\hat \Phi}_3^1={\hat S}_3^1$, 
if $\ket{m}$ is of the form 
\begin{equation}\label{7-14}
\ket{m}=({\hat b}_1(3)^*)^{m_1}\ket{0}=
({\hat b}_3^{1*}{\hat b}_2^{2*}-{\hat b}_3^{2*}{\hat b}_2^{1*})^{m_1}
\ket{0} \ . 
\end{equation}
In this case, we have four excited state generating operators, and then, 
this case is essentially in non-symmetric representation.

One of the present authors (M. Y.), Kuriyama and Kunihiro have already 
investigated the case based on the form (\ref{7-14}).\cite{3} 
In this investigation, the following operators are defined: 
\begin{subequations}\label{7-15}
\begin{eqnarray}
& &{\hat D}_+^*={\hat S}^3 \ , \qquad 
{\hat D}_-^*={\hat S}_2^1 \ , \qquad
{\hat D}_0^*=(1/2)({\hat S}^2-{\hat S}_3^1) \ , \nonumber\\
& &{\hat D}_+ ={\hat S}_3 \ , \qquad 
{\hat D}_-={\hat S}_1^2 \ , \qquad
{\hat D}_0=(1/2)({\hat S}_2-{\hat S}_1^3) \ , 
\label{7-15a}\\
& &{\hat M}_+={\hat S}^2+{\hat S}_2^1 \ , \quad 
{\hat M}_-={\hat S}_2+{\hat S}_1^2 \ , \quad 
{\hat M}_0=(1/2)({\hat S}_3^3+{\hat S}_2^2-{\hat S}_1^1) \ , 
\label{7-15b}\\
& &{\hat I}_+={\hat S}^1+{\hat S}_3^2 \ , \quad 
{\hat I}_-={\hat S}_1+{\hat S}_2^3 \ , \quad 
{\hat I}_0=(1/2)({\hat S}_3^3-{\hat S}_2^2+{\hat S}_1^1) \ , 
\label{7-15c}\\
& &{\hat K}_+={\hat S}^1-{\hat S}_3^2 \ , \quad 
{\hat K}_-={\hat S}_1-{\hat S}_2^3 \ , \quad 
{\hat K}_0=(1/2)(-{\hat S}_3^3+{\hat S}_2^2+{\hat S}_1^1) \ . \qquad
\label{7-15d}
\end{eqnarray}
\end{subequations}
In this form, ${\hat D}_{\pm,0}^*$ and ${\hat M}_+$ play a role of the 
excited state generating operators, which are functions of 
${\hat S}^2$, ${\hat S}^3$, ${\hat S}_2^1$ and ${\hat S}_3^1$. 
Further, in Ref.\citen{3}, the state (\ref{7-14}) was already discussed. 
The reason why the form (\ref{7-15}) is investigated is as follows: 
As was mentioned in \S 1, we know two $su(4)$-algebraic model for 
many-fermion systems. 
One is many-nucleon system under the isospin vector and scalar pairing 
correlations. 
Let us consider a single-orbit shell model, in which the 
single-particle state is 
specified by $(\lambda, \mu$). 
Here, $\lambda$ is half-integer and 
$\mu=-\lambda, -\lambda+1, \cdots , \lambda-1, \lambda$. 
In the above shell model, we define the following operators: 
\begin{subequations}\label{7-16}
\begin{eqnarray}
& &{\wtilde D}_+^*
=(1/2)\sum_{\mu}(-)^{\lambda-\mu}{\tilde n}_\mu^*{\tilde n}_{-\mu}^* \ , 
\qquad 
{\wtilde D}_-^*
=(1/2)\sum_{\mu}(-)^{\lambda-\mu}{\tilde p}_\mu^*{\tilde p}_{-\mu}^* \ , 
\nonumber\\
& &{\wtilde D}_0^*
=(1/2)\sum_{\mu}(-)^{\lambda-\mu}{\tilde n}_\mu^*{\tilde p}_{-\mu}^* \ , 
\nonumber\\
& &{\wtilde D}_+ 
=(1/2)\sum_{\mu}(-)^{\lambda-\mu}{\tilde n}_{-\mu}{\tilde n}_{\mu} \ , 
\qquad 
{\wtilde D}_-
=(1/2)\sum_{\mu}(-)^{\lambda-\mu}{\tilde p}_{-\mu}{\tilde p}_{\mu} \ , 
\nonumber\\
& &{\wtilde D}_0
=(1/2)\sum_{\mu}(-)^{\lambda-\mu}{\tilde p}_{-\mu}{\tilde n}_{\mu} \ , 
\label{7-16a}\\
& &{\wtilde M}_+
=\sum_{\mu}(-)^{\lambda-|\mu|}{\tilde n}_\mu^*{\tilde p}_{-\mu}^* \ , 
\qquad 
{\wtilde M}_-
=\sum_{\mu}(-)^{\lambda-|\mu|}{\tilde p}_{-\mu}{\tilde n}_{\mu} \ , 
\nonumber\\
& &{\wtilde M}_0
=(1/2)\sum_{\mu}({\tilde n}_\mu^*{\tilde n}_\mu+{\tilde p}_{\mu}^*
{\tilde p}_\mu)-(2\lambda+1)/2 \ , 
\label{7-16b}\\
& &{\wtilde I}_+
=\sum_{\mu}{\tilde n}_\mu^*{\tilde p}_{\mu} \ , 
\qquad 
{\wtilde I}_-
=\sum_{\mu}{\tilde p}_{\mu}^*{\tilde n}_{\mu} \ , 
\nonumber\\
& &{\wtilde I}_0
=(1/2)\sum_{\mu}({\tilde n}_\mu^*{\tilde n}_\mu-{\tilde p}_{\mu}^*
{\tilde p}_\mu) \ , 
\label{7-16c}\\
& &{\wtilde K}_+
=\sum_{\mu}(\mu/|\mu|){\tilde n}_\mu^*{\tilde p}_{\mu} \ , 
\qquad 
{\wtilde K}_-
=\sum_{\mu}(\mu/|\mu|){\tilde p}_{\mu}^*{\tilde n}_{\mu} \ , 
\nonumber\\
& &{\wtilde K}_0
=(1/2)\sum_{\mu}(\mu/|\mu|)({\tilde n}_\mu^*{\tilde n}_\mu
-{\tilde p}_{\mu}^*{\tilde p}_\mu) \ . 
\label{7-16d}
\end{eqnarray}
\end{subequations}
As is clear from the forms (\ref{7-16a}) and (\ref{7-16b}), 
${\wtilde D}_{\pm,0}^*$ (isovector) and ${\wtilde M}_+$ (isoscalar) 
are identical with nucleon-pair creation operators and they are 
the excited state generating operators. 
Of course, $({\tilde n}_\mu , {\tilde n}_\mu^*)$ and 
$({\tilde p}_\mu , {\tilde p}_{\mu}^*)$ denote the neutron and the proton 
operators for the state $(\lambda, \mu)$. 
The other fermion model is related to the high-temperature 
superconductivity.\cite{5} 
In this model, ${\tilde D}_{\pm,0}^*$ and ${\wtilde D}_{\pm,0}$, 
${\wtilde M}_{\pm,0}$, ${\wtilde K}_{\pm,0}$ and ${\wtilde I}_{\pm,0}$ 
are called the $d$-wave triplet pairing, the $d$-wave singlet pairing, 
the staggered magnetization and the total spin, respectively, 
in many-electron system. 
It may be interesting to see that the above two many-fermion systems 
are in the non-symmetric representation of the $su(4)$-algebra. 
Thus, we can conclude that not only the symmetric but also non-symmetric 
representation plays an unavoidable role in dynamics of many-fermion 
systems.

\section*{Acknowledgements} 
%
The main part of this work was performed when two of the present authors 
(Y. T. and\break M. Y.) stayed at Coimbra in September of 2004. 
They express their sincere thanks to Professor J. da Provid\^encia, 
co-author of this paper, for his kind invitation and hospitality.

\appendix
\section{Applicability of $\ket{m}$ to the finite temperature effects}

The aim of this Appendix is to give a reply to the question 
if $\ket{m}$ is useful for describing the finite temperature effects or not. 
In order to make the problem simplify, we discuss the case of the 
$su(2)$-algebra. 
The notations are used in the same forms as those shown in \S 7. 

First, we note that the orthogonal set for the $su(2)$-algebra in the 
Schwinger boson representation is specified by the quantum numbers $s$ 
and $s_0$, which is shown in the relation (\ref{7-3}). 
The connection of $s$ to the quantities characterizing the Lipkin model 
is also shown in the relation (\ref{7-4}), especially, in the case 
${\cal N}=\Omega$, (\ref{7-5}). 
Clearly, there does not exist any possibility to add new quantum 
number. 
However, the orthogonal set in the Lipkin model is specified not only 
by $s$ and $s_0$ but also by the quantum number additional to $s$ and $s_0$, 
which we denote $\gamma$: $\kket{\gamma; s,s_0}$. 
For example, the orthogonal set in the case $\Omega=4$ consists of 70 states. 
The intrinsic states are classified as follows: 
1 state for $s=2$, 15 states for $s=1$ and 20 states for $s=0$. 
Therefore, the quantum number $\gamma$ is necessary, and in order to 
investigate the finite temperature effects, all these states must be taken 
into account.

For investigation of the finite temperature effects, the calculation of the 
statistical average of physical quantity is necessary. 
We are now interested in the system following the $su(2)$-algebra. 
Therefore, as any physical quantity, we restrict ourselves to the operator 
which is expressed in terms of the $su(2)$-generator. 
For the operator ${\wtilde O}$, we have 
\begin{eqnarray}\label{a1}
\bbra{\gamma;s,s_0}{\wtilde O}\kket{\gamma';s',s_0'}
&=&\bbra{\gamma;s,s_0}{\wtilde O}\kket{\gamma;s,s_0'}
\delta_{\gamma\gamma'}\delta_{ss'}\nonumber\\
&=&\bra{s,s_0}{\hat O}\ket{s,s_0'}\delta_{\gamma\gamma'}\delta_{ss'} \ . 
\end{eqnarray}
Here, ${\hat O}$ is obtained from ${\wtilde O}$ by replacing 
${\wtilde S}_{\pm,0}$ with ${\hat S}_{\pm,0}$. 
It should be noted that 
$\bbra{\gamma;s,s_0}{\wtilde O}\kket{\gamma';s',s_0'}$ does not depend on 
$\gamma$ (and $\gamma'$) and if $s\neq s'$, it vanishes. 
This fact suggest that the pure states are specified by $\gamma$ and $s$ 
and the state $\kket{\gamma;s}$ is expanded in the form 
\begin{equation}\label{a2}
\kket{\gamma,s}=\sum_{s_0}C_{\gamma;s,s_0}\kket{\gamma;s,s_0} \ . 
\end{equation}
Therefore, the expectation value of ${\wtilde O}$ for 
$\kket{\gamma,s}$ is obtained as follows: 
\begin{equation}\label{a3}
\bbra{\gamma;s}{\wtilde O}\kket{\gamma;s}=\sum_{s_0,s_0'}
C_{\gamma;s,s_0}^*C_{\gamma;s,s_0'}\bra{s,s_0}{\hat O}\ket{s,s_0'} \ . 
\end{equation}
The above gives us the statistical average of ${\wtilde O}$ in the form 
\begin{eqnarray}\label{a4}
\langle{\wtilde O}\rangle_{\rm AV}&=&
\sum_{\gamma s}w_{\gamma;s}\bbra{\gamma;s}{\wtilde O}\kket{\gamma;s}
\nonumber\\
&=&\sum_{ss_0s_0'}\bra{s,s_0}{\hat O}\ket{s,s_0'}\sum_{\gamma}
C_{\gamma;s,s_0'}w_{\gamma;s}C_{\gamma;s,s_0}^* \ . 
\end{eqnarray}
Here, $w_{\gamma;s}$ denotes the statistical weight. 
The form (\ref{a4}) can be rewritten as 
\begin{equation}\label{a5}
\langle{\wtilde O}\rangle_{\rm AV}={\rm Tr}({\hat O}{\hat \rho}) \ . 
\end{equation}
The operator ${\hat \rho}$ denotes the density matrix defined by 
\begin{equation}\label{a6}
\rho_{s,s_0';s,s_0}
=\sum_{\gamma}C_{\gamma;s,s_0'}w_{\gamma;s}C_{\gamma;s,s_0}^* \ . 
\end{equation}
The relation (\ref{a6}) tells us that $\rho_{s,s_0';s,s_0}$ does not depend on 
$\gamma$, and then, for the calculation of 
$\langle{\wtilde O}\rangle_{\rm AV}$, 
explicit knowledge on $\gamma$ is not necessary. 
Further, the matrix element $\bra{s,s_0}{\hat O}\ket{s,s_0'}$ is calculated 
in the framework of the Schwinger boson representation. 
The above argument may be generalized to the Lipkin model in the 
$su(M+1)$-algebra.

\end{document}